\documentclass[aip,rsi,amsmath,amssymb,reprint,]{revtex4-1}
\usepackage{textcomp}
\usepackage{graphicx}
\usepackage{dcolumn}
\usepackage{bm}
\usepackage{siunitx}

\begin{document}

\preprint{AIP/123-QED}

\title[Stockholm, \today]{On-demand generation of background--free single photons from a solid-state source}

\author{Lucas~Schweickert}
\author{Klaus~D.~J\"ons}
\email[Corresponding author: ]{klausj@kth.se}
\author{Katharina~D.~Zeuner}
\affiliation{Department of Applied Physics, Royal Institute of Technology, Albanova University Centre, Roslagstullsbacken 21, 106 91 Stockholm, Sweden}%
\author{Saimon~Filipe~Covre~da~Silva} 
\author{Huiying~Huang}
\affiliation{Institute of Semiconductor and Solid State Physics, Johannes Kepler University Linz, 4040, Austria}%
\author{Thomas Lettner}
\affiliation{Department of Applied Physics, Royal Institute of Technology, Albanova University Centre, Roslagstullsbacken 21, 106 91 Stockholm, Sweden}%
\author{Marcus~Reindl}
\affiliation{Institute of Semiconductor and Solid State Physics, Johannes Kepler University Linz, 4040, Austria}%
\author{Julien Zichi}
\affiliation{Department of Applied Physics, Royal Institute of Technology, Albanova University Centre, Roslagstullsbacken 21, 106 91 Stockholm, Sweden}%
\author{Rinaldo~Trotta}
\affiliation{Institute of Semiconductor and Solid State Physics, Johannes Kepler University Linz, 4040, Austria}%
\affiliation{Dipartimento di Fisica, Sapienza Universit\`a di Roma, Piazzale A. Moro 1, I-00185 Roma, Italy}%
\author{Armando~Rastelli}
\affiliation{Institute of Semiconductor and Solid State Physics, Johannes Kepler University Linz, 4040, Austria}%
\author{Val~Zwiller}
\affiliation{Department of Applied Physics, Royal Institute of Technology, Albanova University Centre, Roslagstullsbacken 21, 106 91 Stockholm, Sweden}%
\affiliation{Kavli Institute of Nanoscience, Delft University of Technology, Lorentzweg 1, 2628CJ Delft, The Netherlands}%

\date{\today}

\begin{abstract}
True on--demand high--repetition--rate single--photon sources are highly sought after for quantum information processing applications.
However, any coherently driven two-level quantum system
suffers from a finite re-excitation probability under pulsed excitation, causing undesirable multi--photon emission. Here, we present a solid--state source of on--demand single photons yielding a raw second--order coherence of $g^{(2)}(0)=(7.5\pm1.6)\times10^{-5}$ without any background subtraction nor data processing. To this date, this is the lowest value of $g^{(2)}(0)$ reported for any single--photon source even compared to the previously best background subtracted values. We achieve this result on GaAs/AlGaAs quantum dots embedded in a low--Q planar cavity by employing (i) a two--photon excitation process and (ii) a filtering and detection setup featuring two superconducting single--photon detectors with ultralow dark-count rates of $(0.0056\pm0.0007)$\,\si{\per\second} and $(0.017\pm0.001)$\,\si{\per\second}, respectively. Re--excitation processes are dramatically suppressed by (i), while (ii) removes false coincidences resulting in a negligibly low noise floor.
\end{abstract}

\pacs{Valid PACS appear here}
\maketitle

Recently, scientific and industrial interest in quantum simulation, computation and communication applications has shown considerable increase.~\cite{Riedel.Binosi.ea:2017} In the field of quantum information processing and communication~\cite{Zoller.Beth.ea:2005} single photons have emerged as ideal candidates for quantum information carriers (flying qubits) due to their small interaction cross--section. Applications with particularly stringent requirements on the second--order coherence are current protocols in cluster-state computation and the realization of an all-optical quantum repeater.~\cite{Aharonovich.Englund.ea:2016}
Like fiber amplifiers in classical optical long--distance communication, quantum repeaters have to be employed to increase the range over which a quantum channel can reliably function. A lower multi--photon emission rate allows for a higher number of consecutive repeater nodes without negatively affecting the secret--key rate, therefore allowing longer distance communication.~\cite{Guha.Krovi.ea:2015,Khalique.Sanders:2015}
Current single--photon sources include trapped atoms, heralded spontaneous parametric down conversion sources, color centers and the emerging field of 2D materials, as well as semiconductor quantum dots (QDs).~\cite{Aharonovich.Englund.ea:2016} Spontaneous parametric down conversion sources offer room temperature operation, but suffer from intrinsic multi--photon emission scaling with the emission rate.~\cite{Waks.Diamanti.ea:2006} The lowest second--order coherence at time delay zero demonstrated with natural atoms is $g^{(2)}(0)=(3\pm1.5)\times10^{-4}$ with background subtraction.~\cite{Higginbottom.Slodicka.ea:2016}  However, these systems typically suffer from a low repetition rate, limited by their intrinsically long lifetime. Compared to natural atoms and ions, optically active semiconductor quantum dots are scalable, nano--fabricated, high repetition rate single--photon sources with tailorable optical properties.~\cite{Michler:2017} QDs under direct resonant excitation have only shown $g^{(2)}(0)=(2.8\pm1.2)\times10^{-3}$ with background subtraction.~\cite{Somaschi.Giesz.ea:2016}
The reason for this is that a residual multi--photon emission probability cannot be fully suppressed under direct resonant pulsed excitation of the excited state (exciton (X) or charged exciton).~\cite{Dada.Santana.ea:2016, Fischer.Muller.ea:2016, Fischer.Hanschke.ea:2017} This holds true for all quantum mechanical two--level systems. Addressing a QD via a third level should result in even lower multi--photon emission probability due to suppressed re--excitation processes. However, this has only been demonstrated using temporal post--selection of coincidence events, yielding $g^{(2)}(0)=(4.4\pm0.2)\times10^{-4}$.~\cite{Miyazawa.Takemoto.ea:2016} In this work, we employ two--photon resonant excitation of the biexciton (XX) state~\cite{Jayakumar.Predojevic.ea:2013,Muller.Bounouar.ea:2014}, strongly suppressing multi--photon emission of our quantum dot and thereby reaching an unprecedented second--order coherence of $g^{(2)}(0)=(7.5\pm1.6)\times10^{-5}$.\\
The QD sample was grown by molecular beam epitaxy at Johannes Kepler University Linz. The QD layer is obtained by Al-droplet etching~\cite{Heyn.Stemmann.ea:2009,Huo.Witek.ea:2013} on Al$_{0.4}$Ga$_{0.6}$As followed by deposition of \SI{2}{\nano\meter} GaAs. This technique allows the fabrication of highly symmetric QDs with measured entanglement fidelities of up to \SI{94}{\percent}.~\cite{Huber.Reindl.ea:2017} The QD layer is placed at the center of a $\lambda$-cavity made of a $\lambda$/2-thick (\SI{123}{\nano\meter}) layer of Al$_{0.4}$Ga$_{0.6}$As sandwiched between two $\lambda$/4-thick (\SI{59.8}{\nano\meter}) Al$_{0.2}$Ga$_{0.8}$As layers. The cavity sits on top of a distributed Bragg reflector made of 9 pairs of  $\lambda$/4-thick Al$_{0.95}$Ga$_{0.05}$As (\SI{68.9}{\nano\meter}) and Al$_{0.2}$Ga$_{0.8}$As layers and below two pairs of the same material combination. A 4 nm-thick GaAs protective layer completes the structure. The QD emission is centered around $\sim$\,\SI{790}{\nano\meter} and a gradient in the mode position (Q factor of about 50) is obtained by stopping the substrate rotation during the deposition of the top Al$_{0.2}$Ga$_{0.8}$As cavity-layer. This simple cavity design enhances the extraction efficiency by $\sim$\,15 times compared to an unstructured sample.
\begin{figure}[htb]
\includegraphics[width=\columnwidth]{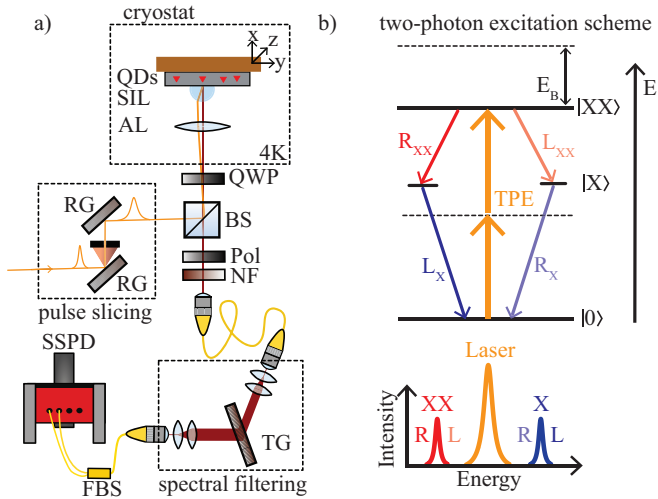}
\caption{\label{fig:setup} a) Confocal micro--photoluminescence spectroscopy setup with pulse--slicer, closed--cycle cryostat, polarization suppression, transmission spectrometer and superconducting single--photon detectors (SSPD). Additional optical components: reflection grating (RG), polarizer (Pol), beamsplitter (BS), quarter waveplate (QWP), aspheric lens (AL), solid immersion lens (SIL), notch filter (NF), transmission grating (TG), fiber beamsplitter (FBS). b) Two--photon excitation scheme to resonantly excite the biexciton state with two laser photons (orange). Top: Three level energy scheme of the biexciton--exciton cascade. Bottom: Visualization of the two--photon excitation spectrum.}
\end{figure}
As illustrated in Fig.~\ref{fig:setup}\,a) the sample is cooled to about \SI{4}{\kelvin} in a closed--cycle cryostat. Inside the cryostat we use an aspheric lens with a working distance of \SI{4.93}{\milli\meter} to focus the excitation laser under a slight angle through a solid immersion lens in the Weierstra\ss{} geometry and onto the QD under investigation. \SI{90}{\percent} of the confocally collected photoluminescence then passes the non--polarizing beam splitter we use to couple in the pulse--stretched excitation laser with a repetition rate of \SI{80.028}{MHz}. As shown in Fig.~\ref{fig:setup}\,b), we tune a pulsed laser to an energy corresponding to half the energy difference between the ground state and the biexciton state in order to resonantly address the biexciton state of the QD with a two--photon process.~\cite{Brunner.Abstreiter.ea:1994,Stufler.Machnikowski.ea:2006} In case of the quantum dot under investigation this corresponds to a laser wavelength of \SI{793.8}{\nano\meter}. Using a pulse shaper we create a laser pulse with a spectral width of
\SI{260}{\micro \electronvolt}, measured with a spectrum analyzer and a pulse length of \SI{7}{ps}, measured with an auto--correlator assuming a Gaussian pulse shape. The peak power density of a $\pi$--pulse is \SI{96}{\kilo\watt\per\square\centi\meter} and the scattered laser light is subsequently filtered. Since, in the case of two--photon resonant excitation (see Fig.~\ref{fig:setup}\,b), bottom), the excitation energy is detuned from the emission energies of both XX and X, we can suppress the laser spectrally. After the beam splitter we use tunable notch filters ($\text{FWHM}=0.4$\,\si{\nano\meter}, extinction ratio $\sim$\,\SI{30}{\decibel}) mounted on stepper motors to selectively block laser light before we couple into an optical fiber. The fiber core with a diameter of \SI{4.4}{\micro\meter} acts as a spatial filter. In addition, we employ polarization suppression by cross--polarizing excitation and detection photons.~\cite{Kuhlmann.Houel.ea:2013} Furthermore, a transmission spectrometer with a bandwidth of \SI{22}{\pico\meter} and an end--to--end efficiency of \SI{60}{\percent} suppresses the remaining light at all wavelengths, except the XX photons. Specifically, the laser, spectrally detuned by \SI{0.9}{\nano\meter} with respect to the XX wavelength, is suppressed by \SI{86}{\decibel}. We then use a fiber based 50\,:\,50 beam splitter to send the photons onto two superconducting single--photon detectors (SSPDs), with a FWHM timing jitter of \SI{20}{\pico\second} and \SI{30}{\pico\second} and dark count rates of $(0.0056\pm0.0007)$\,\si{\per\second} and $(0.017\pm0.001)$\,\si{\per\second}, to perform a start--stop measurement. At these settings and for the biexciton wavelength of $\sim$\,\SI{795}{\nano\meter} our detectors still perform with detection efficiencies of \SI{50}{\percent} and \SI{64}{\percent}, respectively.
\begin{figure}[htb]
\includegraphics[width=\columnwidth]{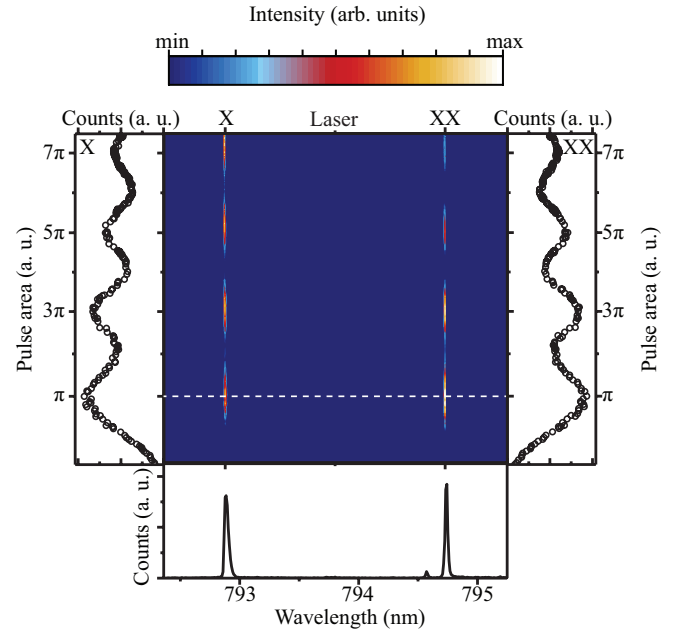}
\caption{\label{fig:rabi} Color--coded photoluminescence spectra of the biexciton--exciton cascade under resonant two--photon excitation as a function of the excitation pulse area. Integrated peak intensity of the X (XX) emission is shown to the left (right) of the color map, undergoing Rabi--oscillations. The spectrum shown in the bottom is excited with a pulse area corresponding to a $\pi$--pulse and is indicated by the horizontal dashed line.}
\end{figure}

In order to verify we are addressing our quantum system coherently via two--photon excitation, we investigate the power dependence of the photoluminescence. Fig.~\ref{fig:rabi} is a color plot showing spectra of the QD including both the X and XX emission for different excitation pulse areas. Blue corresponds to low intensity, whereas red corresponds to high intensities. The observed Rabi--oscillations of the intensity reflect an oscillation in the population of the excited state (XX) and indicate that the system is coherently driven by the excitation light field without the need for any additional off-resonant light field.~\cite{Reindl.Jons.ea:2017} 
To the left (right) of the color plot we show the integrated intensity of the exciton (biexciton) transition as a function of the excitation pulse area. The quadratic power dependence of the state population reflects the two--photon nature of the excitation process, clearly visible in the initial rise.
Below the color plot we show an exemplary spectrum excited with a pulse area corresponding to the maximum population inversion probability ($\pi$--pulse) indicated by the horizontal dashed white line in the color plot. The small peak next to the biexciton has a linear dependence on the laser pulse area and could not be attributed to a specific quantum dot transition.

To investigate the multi--photon emission probability of our source we record coincidences of XX photons between both output ports of a 50\,:\,50 beam splitter binned in time windows of \SI{16}{\pico\second}.
\begin{figure}[htb]
\includegraphics[width=\columnwidth]{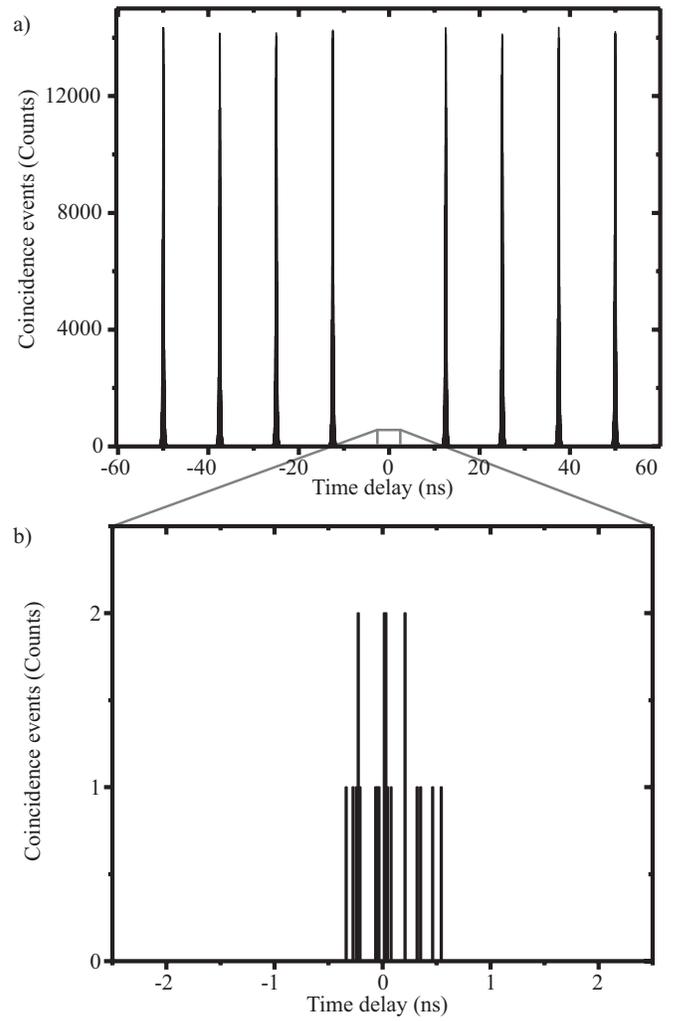}
\caption{\label{fig:g2} a) Measured second--order autocorrelation function of the biexciton under $\pi$-pulse two--photon excitation. An average side peak contains $279171\pm187$ coincidences. b) \SI{5}{\nano\second} wide zoom-in around $\tau=0$, showing a total of $21\pm5$ coincidences. We use this time window for the calculation of the $g^{(2)}(0)$ value.}
\end{figure}
In Fig.~\ref{fig:g2}\,a) we show the resulting histogram approximating a pulsed second--order intensity autocorrelation $g^{(2)}(\tau)$--function. The distance between the side peaks is \SI{12.496}{\nano\second} corresponding to the laser repetition rate. In order to analyze our multi--photon emission probability we compare the amount of coincidences from consecutive excitation pulses with the amount of coincidences within the same pulse. We choose a time window of \SI{5}{\nano\second}, $\sim$\,40 times longer than the XX lifetime of $125\,\si{\pico\second}$ (where the accuracy is limited by the detector time jitter of \SI{20}{\pico\second}) to avoid temporal post--selection. In this time window, we sum up the coincidences to find an average of $279171\pm187$ events per side peak in a sample of 8 side peaks after integrating for \SI{10}{\hour}. The error is propagated quadratically from the square root of the counts in each individual side peak, assuming Poissonian counting statistics. In Fig.~\ref{fig:g2}\,b), we show a \SI{5}{\nano\second} window centered around $\tau=0$ with only $21\pm5$ coincidence events, where the error is the square root of the coincidences, leading to a value of $g^{(2)}(0)=(7.5\pm1.6)\times10^{-5}$. The error is based on the statistical error
and was calculated using quadratic error propagation. 
Given our low $g^{(2)}(0)$ value together with a detected single--photon count rate of $(60\pm5)$\,\si{\kilo cts\per\second} we were able to experimentally verify the quantum non-Gaussian character~\cite{Filip.Mista:2011,Jezek.Straka.ea:2011} of the single photons emitted from our semiconductor quantum dot. We find a non-Gaussian depth of $5.2\pm1.5$\,\si{\decibel} using the expressions of Ref~\onlinecite{Straka.Predojevic.ea:2014}.

We would like to note that cross--polarization of the emission and detection photons help to lower the measured second--oder coherence function at time delay zero from $g^{(2)}(0)=(4.6\pm0.5)\times10^{-4}$ to the stated record value. This is due to finite laser intensity at the XX energy and spatial position of the fiber core. 
Another reason for the low multi--photon emission probability is the previously described two--photon excitation technique directly addressing the transition from the ground state to the XX state. Unlike with resonant excitation of the X state, where re--excitation can directly occur after the initial emission of the single photon, re--excitation is strongly suppressed in the case of two--photon excitation. Re--excitation can only occur once the system has returned to the ground state - a condition that is delayed by the X state's lifetime of $\sim$\,\SI{210}{\pico\second}. By the time the system has completed its cascaded decay the intensity of the excitation laser pulse is much lower than it would be after only a single decay. A theoretical model of this re--excitation suppression has now been developed during the editorial process of our work.\cite{2018arXiv180101672H} In addition, a lowered excitation laser intensity corresponds to a quadratically lowered re-excitation probability, due to the two-photon nature of the excitation.\cite{Kolesov.Xia.ea:2012}

In summary we have shown a single--photon source with unprecedentedly low multi--photon emission. Measuring $g^{(2)}(0)=(7.5\pm1.6)\times10^{-5}$ without any background subtraction nor temporal post--selection was possible due to two key factors: a low dark coincidence count rate of our SSPDs ($0$ events in \SI{48}{\hour}) and a suppressed re--excitation probability during the lifetime of the X made possible by two--photon resonant excitation of the XX state. This highlights semiconductor quantum dots resonantly excited with a two--photon process as ideal candidates for all--optical quantum repeaters, cluster state computation and other applications where low multi--photon emission is of crucial importance.

K.D.J. acknowledges funding from the MARIE SK\L ODOWSKA-CURIE  Individual  Fellowship under REA grant agreement No. 661416 (SiPhoN). K.D.Z. gratefully acknowledges funding by the Dr. Isolde Dietrich Foundation. This work was financially supported by the European Research Council (ERC) under the European Union’s Horizon 2020 research and innovation programme (SPQRel, grant agreement 679183 and NaQuOp, grant agreement 307687), the Swedish Research Council under grant agreement 638-2013-7152, the Austrian Science Fund (FWF): P29603 and the Linnaeus Center in Advanced Optics and Photonics (Adopt). The JKU group acknowledges V. Volobuev, Y. Huo, P. Atkinson, G. Weihs and B. Pressl for fruitful discussions. The Quantum Nano Photonics group at KTH acknowledges the continuous support by the companies APE Angewandte Physik und Elektronik GmbH on their picoEmerald system and Single Quantum BV on their SSPDs. 
\end{document}